\documentclass[iop,cha,twocolumn]{revtex4}
\usepackage{amsfonts,amsmath,amssymb,latexsym,epsfig}

\newcommand{\es}[1]{\begin{equation}\begin{split}#1\end{split}\end{equation}}

\newcommand{\R}{\mathbb{R}}
\newcommand{\N}{\mathbb{N}}

\begin{document}
\title{Quantifying intermittency in the open drivebelt billiard}
\author{Carl P. Dettmann$^{1}$ and Orestis Georgiou$^{2}$}
\affiliation{1 School of Mathematics, University of Bristol, Bristol BS8 1TW, UK.}
\affiliation{2 Max-Planck-Institute for the Physics of Complex Systems, 01187, Dresden, Germany.}
\begin{abstract}
A ``drivebelt'' stadium billiard with boundary consisting of
circular arcs of differing radius connected by their common tangents
shares many properties with the conventional ``straight''
stadium, including hyperbolicity and mixing, as well as intermittency due to marginally
unstable periodic orbits (MUPOs). Interestingly, the roles of the straight and
curved sides are reversed. Here we discuss intermittent properties of the chaotic trajectories from the
point of view of escape through a hole in the
billiard, giving the exact leading order coefficient $\lim_{t\to\infty} tP(t)$ 
of the survival probability $P(t)$ which is algebraic for fixed hole size. However,
in the natural scaling limit of small hole size inversely proportional to time, the
decay remains exponential.  The big distinction between the straight and drivebelt
stadia is that in the drivebelt case there are multiple families of MUPOs leading to
qualitatively new effects. A further difference is that most  marginal periodic orbits
in this system are oblique to the boundary, thus permitting applications that
utilise total internal reflection such as microlasers.
\end{abstract}

\maketitle
{\bf
Intermittency is a general dynamical phenomenon in which a single chaotic trajectory is
interrupted at apparently random intervals by quasi-regular behavior of varying and
unbounded duration.  While first described in the context of turbulence, it  is ubiquitous
in nature since one of the sources of intermittency, elliptic islands approached
more and more closely by the long quasi-regular intervals, are generic features of
Hamiltonian systems.  Another, and more mathematically tractable, source of intermittency
is that of marginally unstable periodic orbits (MUPOs).  In contrast to elliptic islands,
MUPOs can exist in systems with strong ergodic properties such as the celebrated
stadium billiard introduced by Bunimovich in 1974.  This billiard consists of a particle
confined to a region bordered by two straight parallel lines and two semicircles tangent
to these, in which the MUPOs are ``bouncing ball'' orbits reflecting between the parallel
sides indefinitely.  Orbits with many consecutive bounces on the straight segments are
quasi-regular, and when reaching a curved end of the billiard, they return to make
a quasi-regular sequence of similar length; only after many such sequences does the
orbit return to the chaotic region of phase space.  In the case of two circular arcs connected
by non-parallel tangents, the drivebelt stadium, orbits between the straight segments
are of bounded length, but orbits close to periodic orbits in the larger arc are now
quasi-regular.  After slowly precessing and reaching a straight segment, an analogous
return takes place in a sequence of similar length in the reverse direction.  Now, however,
there are potentially many such families of MUPOs, leading to the richer phenomena of
multiple intermittency.  Here we quantify this multiple intermittency by drilling a hole in
the billiard and waiting for the particle to escape.  The probability of surviving for a specified
time, $P(t)$ decays algebraically, with a coefficient expressed exactly as a sum of contributions
from the MUPOs; numerical simulations confirm these calculations.  Properties of the
straight stadium are retained, but new possibilities arise from the multiple intermittency,
which resembles more closely a generic mixed system with many elliptic islands.
Applications in atom, electron and optical billiards are discussed; unlike
the straight stadium, the oblique reflections of typical MUPOs in the drivebelt now allow
total internal reflection, which allows trapping at the very small scales needed for
microlasers.  These wave billiards also raise intriguing questions for the future - how is
multiple intermittency manifest in the quantum regime?     
}

\section{Introduction}
The study of mathematical billiards \cite{Sinai04} has received much attention in recent years as it facilitates a better understanding of the underlying dynamics of many physical systems while also providing useful models with applications in areas ranging from nonlinear dynamics and statistical physics to quantum optics. Classically, the billiard's phase space is defined by the position and momentum variables of a point particle moving with constant speed in a straight line inside a compact domain $Q\in\R^d$ experiencing mirror-like reflections at the billiard boundary $\partial Q$. As the particle trajectories do not depend on the overall energy, the billiard dynamics (integrable, mixed or chaotic) is completely controlled by the boundary geometry. The addition of a ``hole" or ``leak" in phase space through which an initial ensemble of particles may escape through was first suggested by Pianigiani and Yorke \cite{Yorke80}. This made the survival probability $P(t)$ (a monotonically decreasing function of time $t$) an important statistical observable, able to describe and classify the internal dynamics of the billiard or other dynamical systems. Moreover, it motivates the following question: {\em How do the long-time escape properties depend on the dynamics, the size and position(s) of the hole(s)?}

To a first approximation, the exponential rate of escape in strongly chaotic dynamical systems is proportional to the size of the hole, inversely
proportional to the size of the accessible part of the phase space and independent of the hole's position. We shall be concerned with two dimensional billiards where this statement can be simply expressed as
\begin{equation}\label{e:1}
\gamma=\lim_{t\rightarrow\infty}-\frac{\ln P(t)}{t}= \frac{\epsilon}{\pi |Q|},
\end{equation}
such that $\frac{\pi |Q|}{|\partial Q|}$ is the mean free path between collisions, $\epsilon$ is the hole size, and $|Q|$ and $|\partial Q|$ are the area and perimeter of the billiard respectively. Equation (1) is a good approximation for small holes and times longer than the scale set by correlation
decay which is proportional to one over the Lyapunov exponent of the chaotic saddle. Exact formulas for the subsequent correction terms were derived recently in \cite{DettBun07} as sums of correlation functions depending on both hole size and position, and were numerically tested in the dispersing diamond billiard with a hole placed in coordinate space along the billiard boundary. Holes in momentum space have also been considered and are of interest when studying for example emission from dielectric micro-cavities \cite{Gmachl98}.
The full escape properties of open systems are in theory described by the properties of the invariant chaotic saddle(s) along which a conditionally invariant measure concentrates in time \cite{DemersYoung06} but are however difficult to compute in practice \cite{Altmann09}.
Nevertheless, although placing holes through which particles may escape was originally a perturbative approach for
generating transient chaos \cite{Yorke80,TelLai08}, it has also been successful as a kind of chaotic spectroscopy \cite{Uzy92} used for retrieving information about the internal dynamics \cite{DettBun05,DettBun07}.

Generic dynamical systems however such as the standard map \cite{Chirikov} are weakly chaotic and have a mixed phase space where Kolmogorov-Arnold-Moser (KAM) hierarchical islands coexist with a chaotic sea. Although the dynamics in the hyperbolic regions of the phase space is exponentially mixing, trajectories approaching the islands `stick' there for arbitrarily long periods of time before they are re-injected back into the `deep' hyperbolic regions of the chaotic sea. The sticky dynamics \cite{Karney83} described above is closely associated with the more general phenomenon of intermittency and is captured by a cross-over from exponential to an asymptotic power-law decay in the survival probability $P(t)\sim t^{-\alpha}$, with $\alpha\geq 1$, and is independent of the hole's position, assuming that it is placed well inside the chaotic part of the phase space. In recent studies \cite{Altman09}, this was interpreted as an effective splitting of the chaotic saddle into hyperbolic and nonhyperbolic components. The question of a universal power-law exponent in mixed phase space systems is still an open problem though recent results suggest $\alpha\approx 0.57$ \cite{Ketz08}.

Power-law asymptotic decays are not restricted to mixed phase space systems. For example, power-law tails of order $\sim t^{-1}$ were observed in the famous Bunimovich stadium billiard \cite{Vivaldi83}, a paradigmatic example of chaotic billiards. The stadium billiard \cite{Bu79} is composed of two semicircles joined smoothly $(C^{1})$ by two parallel straight lines. The defocusing mechanism of chaos present guarantees a positive Lyapunov exponent (exponential separation rate of nearby trajectories) almost everywhere, the exception being a zero-measure continuous family of period two marginally unstable periodic orbits (MUPOs) called ``bouncing ball'' orbits trapped forever between the stadium's parallel walls. Bouncing ball orbits have been shown to lead to an intermittent, quasi-regular behavior which effectively causes the closed stadium to display some weaker chaotic properties such as an algebraic decay of correlations \cite{Zhang05} and a non-standard central limit theorem (CLT) requiring a $t \log t$ normalization \cite{Gouzel,Balint08}. Note that ``whispering gallery'' orbits which slide almost tangentially along the curved boundaries are only intermittent in discrete time but not in continuous time as their maximum length is bounded. Quantum mechanically bouncing ball orbits are known to cause scarring \cite{Gabriel}, deviations from Random Matrix Theory (RMT) (especially in the $\Delta_{3}$-statistics) if not treated appropriately \cite{Sieber93,Graf92}, while the system is not quantum uniquely ergodic \cite{Hassel10} and an $\hbar$ dependent `island of stability' appears to surround them \cite{GTan}. Therefore MUPOs are important in understanding the quantum to classical correspondence \cite{Bohr} of the phenomenon of stickiness.

It was recently found by the authors that by positioning a hole on the boundary as to intersect the continuous family of bouncing ball orbits it is possible to explicitly single out long surviving orbits hence calculating the constant $\mathcal{C}$ in closed form as a function of hole position \cite{OreDet09} hence supporting the interpretation of Ref.\cite{Altman09}. Unlike past investigations of open stadia \cite{Arm04}, this result included the exact coefficient of the
survival probability, allowing for the accurate prediction and hence optimization of escape time distributions while also providing answers to the aforementioned motivational question. In this paper we review and generalize the approach of \cite{OreDet09} to circular type MUPOs in the context of the \textit{drivebelt} billiard \cite{Kaplan01,Zhang05,Zhang08,Sanders10b}.

\begin{figure}[h]
\begin{center}
\includegraphics[scale=0.4]{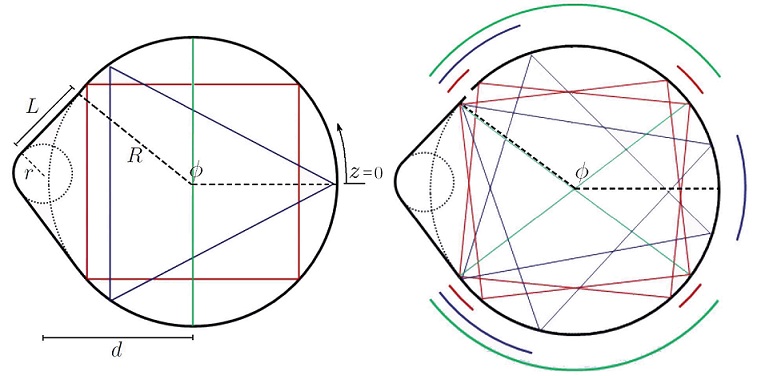}
\caption{\label{fig:drives}\footnotesize The drivebelt billiard is defined by the ratio of the small and big radii $\frac{r}{R}\in(0,1)$ and the arc of the larger of the two circular arcs $\phi\in(\pi/2,\pi)$ which in turns defines which MUPOs exist in the drivebelt's phase space. \emph{Left:} Drivebelt with $\frac{r}{R}=0.3$ and $\phi=2.5$. The period two diametrical MUPO $(s,j)=(2,1)$, the period three triangle MUPO $(s,j)=(3,1)$ and the period four square MUPO $(s,j)=(4,1)$ are shown. \emph{Right:} These MUPOs can be oriented within a certain range depending on the value of $\phi$. A small hole is placed near the edge of the larger arc as to intersect the range of all $3$ MUPOs present.}
\end{center}
\end{figure}

The drivebelt billiard is constructed by two circles of radii $r$ and $R>r$ with centers displaced by a distance $d$. The circles are then connected by their common outer tangent lines of length $L$ (see Figure ~\ref{fig:drives}) and the interior arcs are removed such that the boundary is $C^{1}$ smooth and convex and $d=\sqrt{L^{2}+(R-r)^{2}}$. For the purpose of our current investigation, only half the arc length of the larger circle $\phi\in(\pi/2,\pi)$ needs to be known as shown in Figure ~\ref{fig:drives}. Note that in the limits $\phi\rightarrow \pi/2$ and $\phi\rightarrow \pi$ the stadium billiard and circle billiards are recovered respectively.

This construction, also known as a `tilted' stadium, or a `squash' was originally proposed by Bunimovich (unpublished), investigated numerically and experimentally using ultra cold atoms confined by lasers beams in \cite{Kaplan01}, and later rediscovered in \cite{Zhang05} and \cite{Zhang08}, where it was first called a `drivebelt' billiard and its polynomial mixing rates were studied in detail. Recently, B\'{a}lint \textit{et al.} have studied a two-parameter set of two-dimensional billiards with one of the limiting cases being the drivebelt and conjecture that
they obtain ergodic non-dispersing billiards which are close to their drivebelt limit \cite{Sanders10b}.
The billiard is hyperbolic, ergodic, and Bernoulli \cite{Bu79} and remains chaotic no matter how short $L$ is, but turns into the integrable circle when $L\rightarrow0$. The mechanism of chaos present is the defocusing one (same as for the stadium billiard) where a parallel beam of rays becomes convergent and then divergent after traveling twice the focal length. Therefore expansion in phase space is guaranteed almost everywhere \cite{Wojt86}, the exception here being a set of zero-measure circle-type MUPOs.

Circular type MUPOs (defined later) in a sense generalize bouncing ball orbits as they allow for periods $s\geq2$. They have also been observed in the mixed phase space annular \cite{Saito82} and mushroom billiards \cite{Bun01}. In fact, in both of these cases an infinite number of them was shown to exist \cite{Altman08} while careful investigations involving continued fractions and results from number theory have revealed a MUPO-free set of mushrooms \cite{OreDett10} which can also be extended to the annular billiard. In contrast, the drivebelt billiard has only a finite number of MUPOs controlled by the parameter $\phi$, hence motivating the term \textit{multiple intermittency}. That is, a finite number of regions of the billiard phase space individually display quasi-regular behavior for long periods of time. Interestingly and unlike bouncing ball orbits, circle type MUPOs may have large incidence angles with the boundary and therefore would be interesting to study in the context of micro-resonators \cite{Ansersen09} where total internal reflection will trap some of them according to Snell's law of refraction.
Regardless of this, all aforementioned classical and quantum `side-effects' due to bouncing ball orbits including nonstandard CLT of \cite{Gouzel,Balint08} are expected to be inherited by the drivebelt. Moreover, an interesting duality exists between the stadium's bouncing ball orbits and the drivebelt's circular MUPOs as the roles of straight and curved boundary segments are reversed. Also, as we shall show, they exhibit very similar reflection laws at the end of each quasi-periodic intermittent sequence of reflections, a key property in our approach for calculating the asymptotic survival probability.

In the next section we formally present our model by defining the relevant parameters and briefly discussing how MUPOs affect the open and closed billiard dynamics. In section III, we numerically investigate the open drivebelt phase space and observe the multiple intermittency manifested by the multiple sticky regions surrounding the system's MUPOs. In section IV we adapt and extend the approach of Ref \cite{OreDet09} and obtain an exact to leading order in $t$ analytic expression for the asymptotic survival probability function which we numerically verify and discuss different scaling limits. In section V we conclude with a short discussion.

\section{Circle-type MUPOs and multiple intermittency in the drivebelt billiard}

We define Birkhoff coordinates $(z,\sin\theta)$ such that $z\in[0,|\partial Q|)$ increases from zero anticlockwise from the rightmost point of the billiard as shown in Figure ~\ref{fig:drives} and $\theta\in(-\frac{\pi}{2},\frac{\pi}{2})$ is the angle of incidence. These coordinates are volume preserving under the billiard map and are also the ergodic equilibrium measure of the closed billiard and corresponding volume element $d\mu=(2|\partial Q|)^{-1}\cos\theta\ d\theta d z$ which we use as initial particle ensemble in our numerical simulations since the hole will typically be of small but finite size.

\begin{figure}[h]
\begin{center}
\fbox{
\includegraphics[scale=0.6]{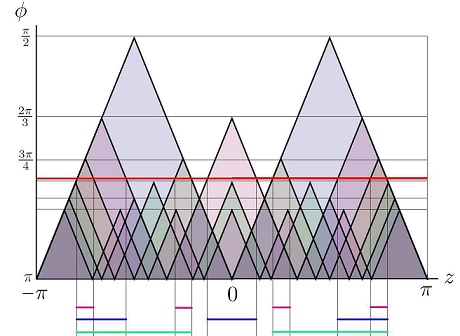}}
\caption{\label{fig:mounts}\footnotesize As $\phi\in(\pi/2, \pi)$ is increased, more and more MUPOs come into existence. Only the first $7$ MUPOs are shown here. The total range they cover also increases monotonically with $\phi$ and is shown on the horizontal axes. The range covered by MUPOs for the example case of $\phi=2.5$ is explicitly shown (also see right panel of Figure ~\ref{fig:drives}).}
\end{center}
\end{figure}
The drivebelt billiard is a non-uniformly hyperbolic system due to the existence of circle-type MUPOs in the larger of its two circular arcs.
MUPOs in the drivebelt are a subset of periodic orbits of the circle billiard and hence have incidence angles given by $\theta_{s,j}=\frac{\pi}{2}-\frac{j \pi}{s}$, where $s\geq2$ and $j\geq1$ are coprime integers describing the period and rotation number respectively. The number of MUPOs in the drivebelt grows approximately as $\sim3/(2(\pi-\phi)^{2})$, hence we coin the term `multiple intermittency'. This is because a MUPO of period $s$ will exist only if it satisfies $\frac{2\pi}{s}>2(\pi-\phi)$ and the number of integers coprime to $s$ goes like $\sim 6s/\pi^2$.
We therefore define the set containing all MUPOs for a given $\phi$ by $\mathcal{S}_{\phi}:=\big\{(2,1), (s,j)\big|3\leq s<\frac{\pi}{\pi-\phi}$, $1\leq j\leq \left\lfloor\frac{s-1}{2}\right\rfloor , \mathrm{gcd}(s,j)=1\big\}$.

The mountain-plot in Figure ~\ref{fig:mounts} shows how MUPOs come into existence as $\phi$ is increased from $\frac{\pi}{2}$ to $\pi$ from top to bottom. Also in Figure ~\ref{fig:mounts} and the right panel of Figure ~\ref{fig:drives} we can see how the families of MUPOs define a range along which they can be oriented inside the larger of the two circular arcs. The example case shown in the right panel of Figure ~\ref{fig:drives} with $\phi=2.5$ is also highlighted in Figure ~\ref{fig:mounts} as a red horizontal line.

The linear stability (monodromy matrix) eigenvalues of MUPOs are both equal to one hence any small perturbation of a MUPO will initially grow linearly (not exponentially) with time as the orbit precesses in the perturbed direction. Some examples are shown in Figure ~\ref{fig:drives}. MUPOs remain periodic after a small perturbation in $z$ and hence form continuous families while perturbations in $\theta$ will cause them to precess and exhibit long periods of quasi-regular behavior (approximately periodic), before leaving the vicinity of the periodic orbit and exploring the rest of the ergodic phase space uniformly and exponentially fast. This quasi-regular, non-dispersive behavior can cause the formation of scars in quantum billiards and is of particular interest in optical microlasers where total internal reflection can pick out some of the MUPOs while allowing others to cause directional emission \cite{Ansersen09}.

Although MUPOs occupy zero volume in phase space and hence do not affect the overall ergodicity of the system, they govern long time statistical properties of the closed system, such as the rate of mixing (the rate of the decay of correlations) $\mathcal{C}(t)\sim t^{-1}$ \cite{Zhang08} and the Poincar\'{e} recurrence times distribution $\mathcal{Q}(t)\sim t^{-2}$ \cite{Altman05} which is also intimately related \cite{Pikovsky92} with the long time survival probability $P(t)\sim t^{-1}$ \cite{Altman06} of the open system. Furthermore, the exponents of these power-laws appear to be a universal fingerprint of non-uniform hyperbolicity and stickiness, at least for one and two dimensional area preserving maps with sharply divided phase space \cite{Altman06}.

To this end, we note that the large number of consecutive collisions that occur in a MUPO's close vicinity in phase space does not violate Kac's famous lemma \cite{Kac}. That is, the mean recurrence time to a small box surrounding a MUPO equals one over the measure of the box. This is of course nothing more than an artifact of ergodicity in that the long times spent by a trajectory stuck in the vicinity of the MUPOs are on average compensated by an apparent `difficulty' in entering such sticky regions in the first place \cite{Arm04}. Note however that due to the sticky dynamics, the second and higher moments of the recurrence times distribution diverge \cite{Zasl02}.

The long periods of quasi-regular behavior and the ergodic compensation mechanism are particularly interesting to discuss in the multiple intermittency setting of the drivebelt billiard.
This is because by defining conditional return and transmission distributions, one may succeed
in breaking the usual independence between events observed in chaotic systems.  For
example, consider small boxes $A$, $B$, etc. in phase space enclosing the MUPOs.
Mixing of this billiard says that the probability of being in $A$ at some time and $B$ at
a much later time approaches $\mu(A)\mu(B)$.  A priori, it does not say what happens at
closely spaced times, and in particular whether the probability of reaching $B$ rather than
$C$ immediately following $A$ is proportional to their respective measures.  If we require
a given number $n>1$ collisions within each region, the measures are reduced according
to the collision rules where the MUPO reaches the straight segments, which leads to a
dependence on the period of the orbit.  Preliminary investigations (omitted) suggest that
these conditional transmission probabilities are remarkably close to the default
independence prediction, but deviations must be evident at some level, in a way that reveals
deeper structure of the dynamics of this, or indeed other, system with multiple intermittency.
We leave further investigation of this point to future work, but note that such a situation
is particularly relevant to optical microlasers where MUPO resonances can interact
via different mechanisms \cite{Ansersen09,Polliner09,Sieber09,Susumu10} and a
minimum number of collisions inside the cavity is needed to achieve high quality factors.
In the next section we perform a numerical investigation of the open drivebelt phase space
and identify the set of sticky orbits responsible for the power-laws discussed above.

\section{Numerical investigations of drivebelt phase space}

\begin{figure}[h]
\begin{center}
\includegraphics[scale=0.27]{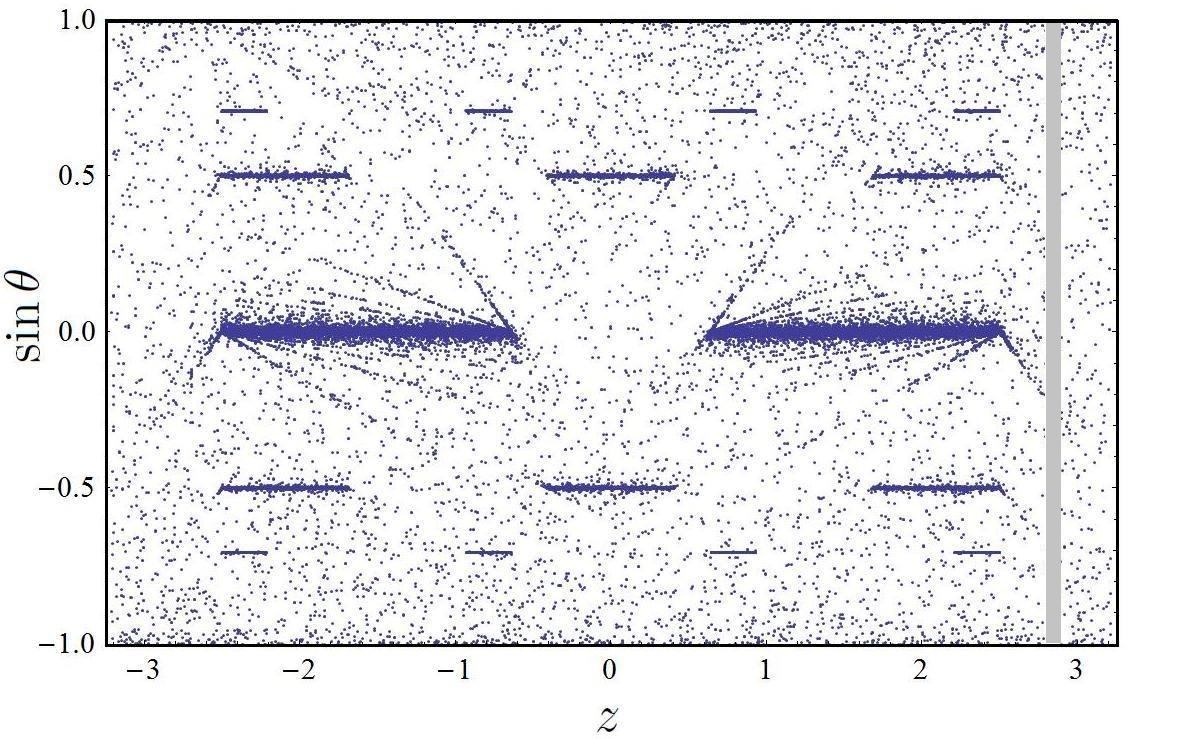}
\caption{\label{fig:PS1}\footnotesize Phase space plot of drivebelt billiard showing $10^{5}$ randomly chosen initial conditions which survive up to time $t=2000$ with drivebelt parameters $\phi=2.5$, $R=1$, $r=0.3$ and a hole of size $\epsilon=0.1$ positioned at $z\in (2.8,2.9)$.}
\end{center}
\end{figure}

We begin our investigations with some phase space images of a drivebelt billiard with $R=1$, $r=0.3$ and $\phi=2.5$. Notice that $\mathcal{S}_{\phi}=\left\{(2,1),(3,1),(4,1)\right\}$. Figure ~\ref{fig:PS1} shows $10^{5}$ randomly chosen initial conditions (ICs) (initially distributed on the billiard boundary according to the equilibrium density) which do not escape after time $t=2000$ through a small hole of size $\epsilon=0.1$ placed at $z\in (2.8,2.9)$. Note that the hole is placed well inside the ergodic component of the phase space but is also away from any sticky regions. The hole can be clearly identified as the thin gray vertical strip on the right of the figure. We notice that long surviving initial conditions seem to populate the sticky regions surrounding the locations of the MUPOs. We also notice that these ICs are supported on spike-like fractal-looking structures which correspond to the unstable manifold of the hole up to $t=2000$. The coarse grained ICs surrounding these spikes correspond to trajectories that move from one MUPO to another but do not intersect the hole up to $t=2000$. Moreover, these long surviving heteroclinic orbits contribute to the asymptotic power-law decay even though they exhibit stable and unstable directions in phase space. 

\begin{figure}[h]
\begin{center}
\includegraphics[scale=0.27]{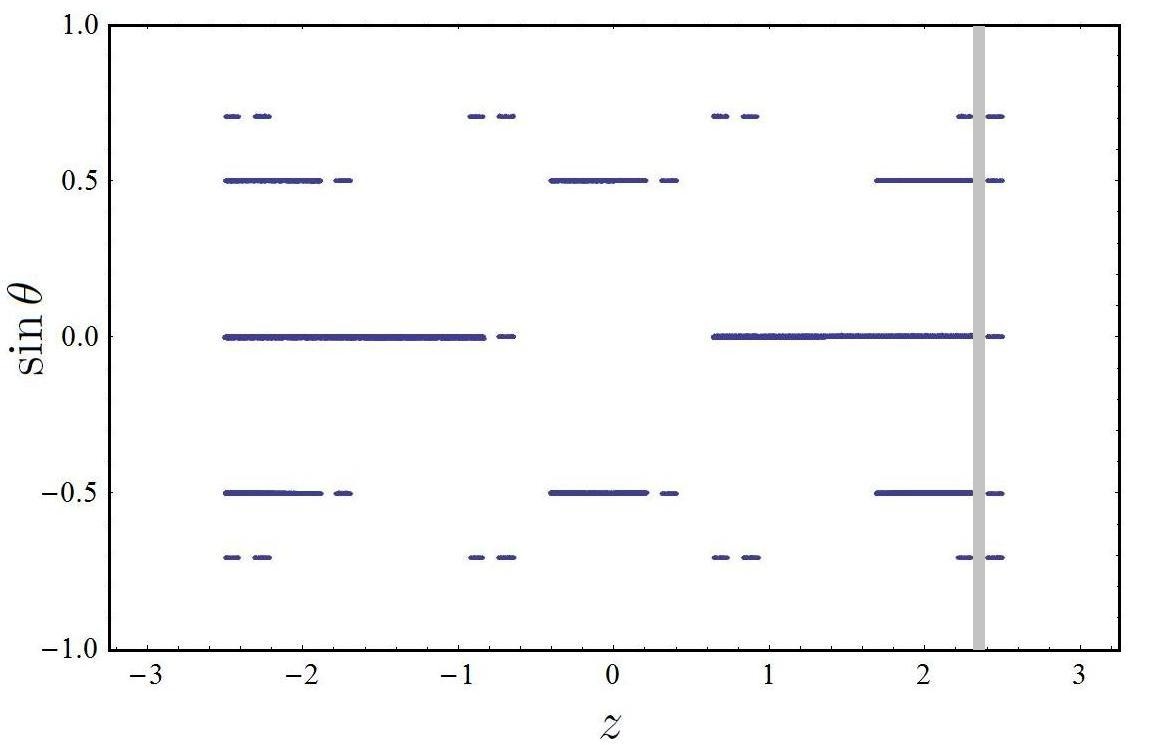}
\caption{\label{fig:PS2}\footnotesize Phase space plot of drivebelt billiard showing $10^{5}$ randomly chosen initial conditions which survive up to time $t=2000$ with drivebelt parameters $\phi=2.5$, $R=1$, $r=0.3$ and a hole of size $\epsilon=0.1$ positioned at $z\in (2.3,2.4)$.}
\end{center}
\end{figure}
\begin{figure}[h]
\begin{center}
\fbox{
\includegraphics[scale=0.23]{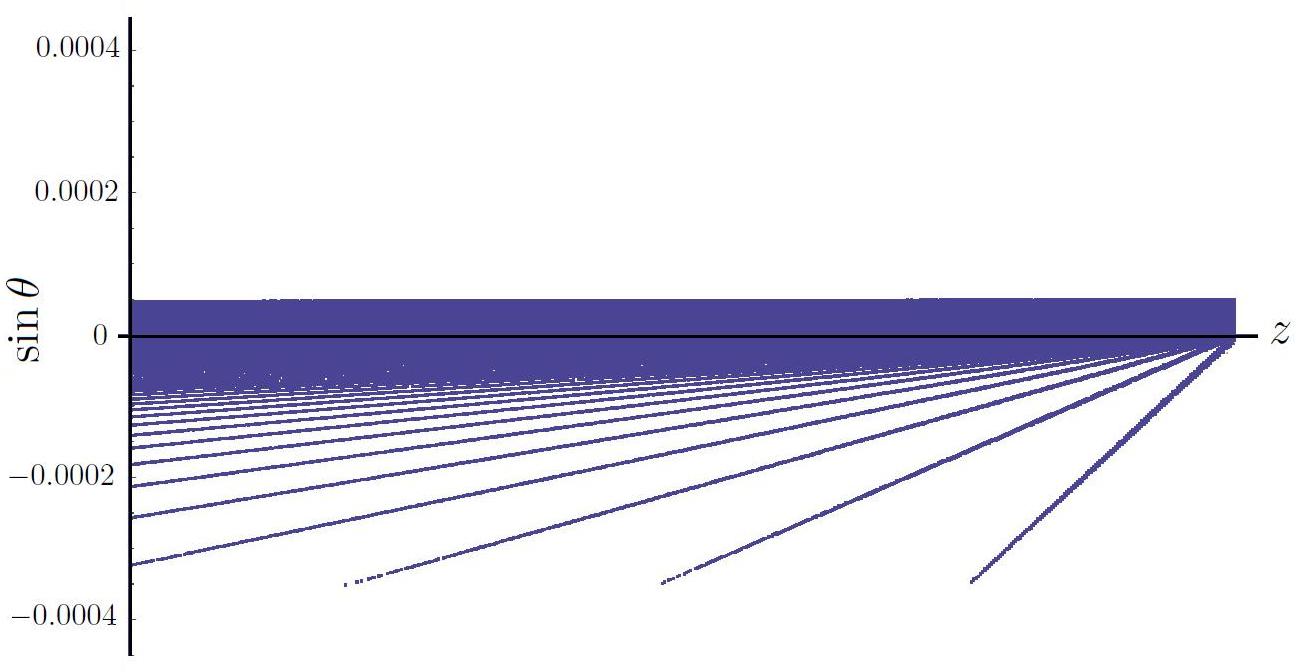}}
\caption{\label{fig:drivespikes}\footnotesize Magnification of phase space plot in Figure ~\ref{fig:PS2} at $z\in(2.45,2.5)$ and $\sin\theta\in(-0.0004,0.0004)$ showing how the long surviving $t>2000$ initial conditions populate the area near the period two diametrical MUPO.}
\end{center}
\end{figure}

Figure ~\ref{fig:PS2} shows $10^{5}$ randomly chosen initial conditions which do not escape after time $t=2000$ through a small hole of size $\epsilon=0.1$ intentionally placed at $z\in (2.3,2.4)$ as to overlap part of all three MUPO ranges. Note that the number of initial conditions plotted in Figures ~\ref{fig:PS1} and ~\ref{fig:PS2} are the same while now the hole clearly repeats itself in all sticky regions. The density of long surviving orbits in regions away from MUPOs appears to have decreased while the spike like structure has shrank dramatically. We therefore zoom in and take a closer look (see Figure ~\ref{fig:drivespikes}). The magnified phase space corresponds to $z\in(2.45,2.5)$ which is right where the larger circular arc meets with the straight. The spike-like fractal-looking structure observed previously is retained for negative $\theta$ but is linear for positive $\theta$. This pattern is preserved in all MUPO vicinities (not shown) suggesting that for long but fixed survival times $t$, small and positive $\theta$ perturbations of MUPOs have a linear dependence with distance to the hole while negative $\theta$ perturbations have a more interesting dependence and can also sustain larger perturbation magnitudes. In the next section we clarify these two types of behaviors at long times by classifying them as ICs precessing towards/away from the hole.

\section{Asymptotic Survival Probability}
Assuming that for large enough times the long surviving initial conditions are dense only in the close vicinity of MUPOs, then a hole which overlaps part of all MUPO ranges will separate the long surviving ICs into two simple families: 1) orbits initially precessing towards the hole, and 2) orbits initially precessing away from the hole. We clarify this classification as follows. A small perturbation $\eta_{i}\ll 1$ in the angle of incidence $\theta_{s,j}$ of a MUPO will cause the orbit to precess in a quasi-periodic fashion. If the orbit is precessing towards the hole and $\eta_{i}$ is small enough then the orbit may survive for an unbounded amount of time and hence will contribute to the asymptotic power-law $\sim t^{-1}$ tail of $P(t)$. In fact such orbits correspond to the ICs with $\theta >0$ shown in Figure ~\ref{fig:drivespikes} and are bounded linearly from above as the precessing velocity is proportional to the perturbation strength. If the orbit is precessing away from the hole and the $\eta_{i}$ is small enough, the orbit will experience a nonlinear collision processes which reverses the orbit's precession direction and is of the type $C\ldots CSC^{s-1}(S)C\ldots C$ until escape. We have used here a symbolic dynamics where the symbols $S$ and $C$ correspond to collisions on straight and curved boundary segments respectively. The $s$ in the superscript corresponds to the period of the nearest MUPO in phase space. Also the $S$ in brackets, as we shall see later only occurs when the first straight segment collision does not reverse the precessing orbit's direction. After the precessing direction is reversed, the orbit will approach the hole with a final angle of incidence which is not far from $\theta_{s,j}$ and may again have unbounded survival time. Such ICs as we shall soon see constitute the spike-like structure observed for $\theta <0$ in Figure ~\ref{fig:drivespikes}. With this classification, we expect that
\es{\label{lim} \lim_{t \rightarrow \infty} tP(t)= \lim_{t \rightarrow \infty} t (P_{1}(t)+P_{2}(t) )}
where $P_{1}(t)$ and $P_{2}(t)$ correspond to the leading order contributions of the two types of ICs.

\subsection{Precessing towards the hole}

Consider a small hole of size $\epsilon$ situated at $z\in (h^{-},h^{+})\subset (0,\phi)$, such that it overlaps all MUPO ranges and $h^{+}=h^{-}+\epsilon$. An example is shown in the right panel of Figure ~\ref{fig:drives}. We perform a change of variables and consider ICs $(\phi_{i},\theta_{i})$ such that $\phi_{i}\in(0, \phi-h^{+})$ is the angular distance of the IC from the nearest edge of the hole in the direction of precession and $\theta_{i}=\theta_{s,j}+\eta_{i}$ is the angle of incidence such that the perturbation $\eta_{i}\ll 1$ causes the orbit to precess towards the hole. Since the precessing angular velocity is proportional to the perturbation strength $\eta_{i}$ the condition
\es{|\eta_{i}|<\frac{\epsilon}{2 s},}
guarantees that the IC does not `jump over' the hole. Hence such an orbit will escape in time $t$ given by
\es{t(\phi_{i},\theta_{i})&= \left\lceil\frac{\phi_{i}}{2\eta_{i}s}\right\rceil s2R\cos\theta_{i}\\
&= \frac{R \phi_{i} \cos\theta_{s,j}}{\eta_{i}}+ \mathcal{O}(1).}
Notice that we have obtained the escape time in a continuous time frame from discrete iterations of the billiard map. This is allowed here since the time between collisions is fixed. From the above information we can obtain a lower bound on the relevant time scales for which our results are valid
\es{|\eta_{i}|< \mathrm{min}\left\{\frac{\epsilon}{2 s}, \frac{R \phi_{i} \cos\theta_{s,j}}{t}\right\},}
and hence when
\es{\label{tt} t>\frac{4\pi R}{\epsilon},}
since $\mathrm{sup}(\phi_{i})=2\pi/s$.

We can now integrate the area described by the above inequalities with respect to the equilibrium measure $(2 |\partial Q|)^{-1}\mathrm{d} z \mathrm{d} \sin \theta = (2 |\partial Q|)^{-1}R(\cos\theta_{s,j}- \eta_{i}\sin\theta_{s,j})\mathrm{d} \phi_{i} \mathrm{d} \eta_{i}$ as follows:
\es{\mathcal{I}_{1}&= \int_{0}^{\frac{\phi_{1}R \cos\theta_{s,j}}{t}} \int_{\frac{t\eta_{i}}{R \cos\theta_{s,j}}}^{\phi_{1}} \frac{R (\cos\theta_{s,j}- \eta_{i}\sin\theta_{s,j})}{2 |\partial Q|}\mathrm{d} \phi_{i} \mathrm{d} \eta_{i}\\
&= \frac{R^{2}\phi_{1}^{2}(3t - R \phi_{1} \sin\theta_{s,j})\cos^{2}\theta_{s,j}}{12 |\partial Q| t^{2}}\\
&= \frac{R^{2}\phi_{1}^{2}\cos^{2}\theta_{s,j}}{4 |\partial Q| t}+ \mathcal{O}\left(\frac{1}{t^{2}}\right) ,}
where $\phi_{1}(s)$ is the angular distance available within the range of the MUPO on the corresponding side of the hole.
Similarly we obtain the asymptotic contribution from ICs precessing towards the hole but from the other side of the hole
\es{\mathcal{I}_{2}= \frac{R^{2}\phi_{2}^{2}\cos^{2}\theta_{s,j}}{4 |\partial Q| t}+ \mathcal{O}\left(\frac{1}{t^{2}}\right) ,}
and the total contribution to the asymptotic survival probability is
\es{P_{1}(t)= \sum_{(s,j)\in \mathcal{S}_{\phi}} s p (\mathcal{I}_{1}+ \mathcal{I}_{2}) +  \mathcal{O}\left(\frac{1}{t^{2}}\right),}
where the $p=1$ if $s=2$ and $p=2$ otherwise as to account for the vertical symmetry of the phase space and $s$ accounts for the multiplicity of the MUPOs as seen in Figures ~\ref{fig:PS1} and ~\ref{fig:PS2}. We next turn to the contribution $P_{2}(t)$ due to ICs precessing away from the hole.

\subsection{Precessing away from the hole}

\begin{figure}[h]
\begin{center}
\fbox{
\includegraphics[scale=0.26]{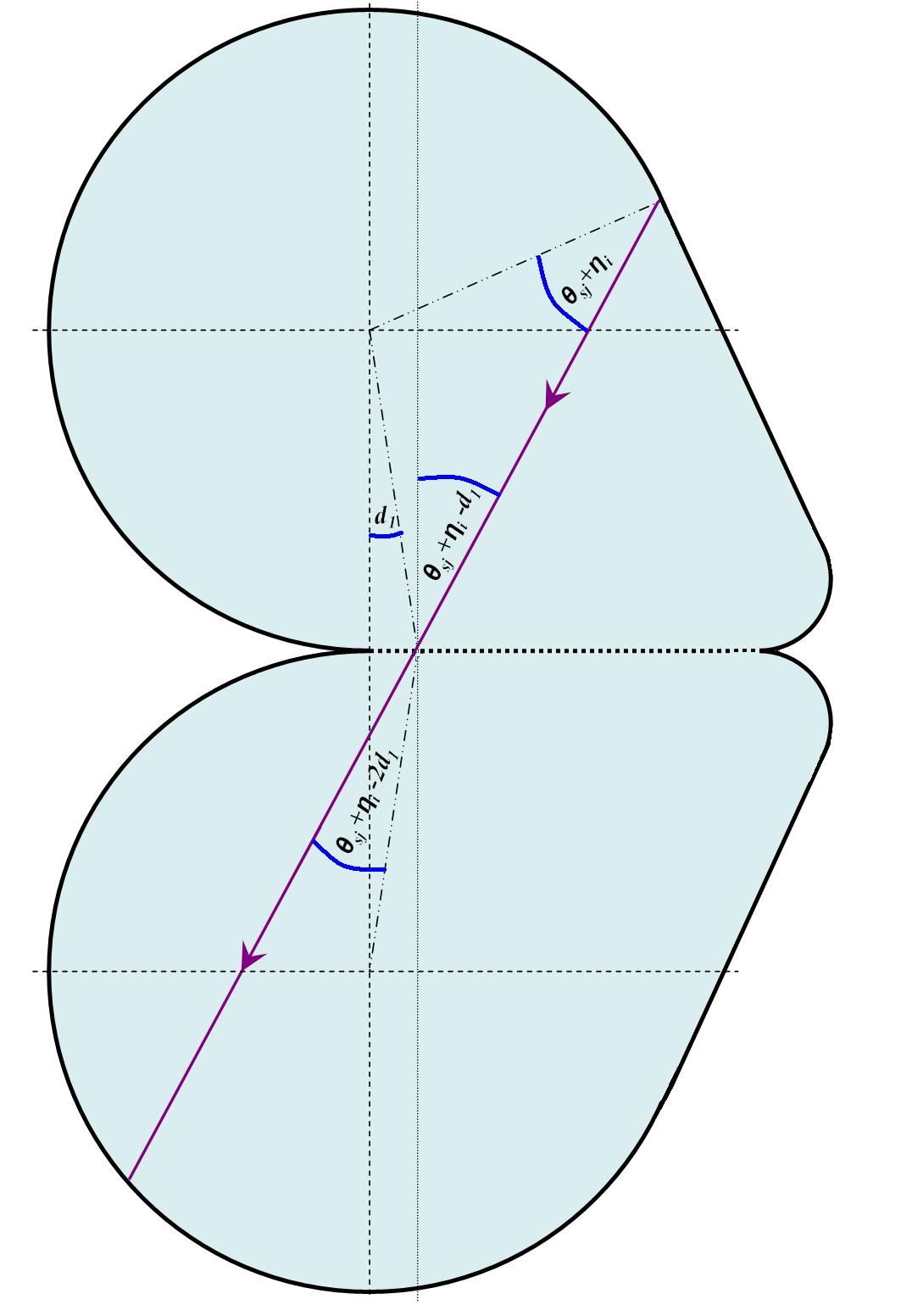}}
\caption{\label{fig:drivebeltcoll}\footnotesize A slightly perturbed in angle MUPO with initial angle of incidence $\theta_{i}=\theta_{sj}+\eta_{i}$ is `unfolded' when it collides on a straight boundary segment giving a new angle of $\theta_{f}=\theta_{i}-2d_{1}=\theta_{s,j}+\eta_{f}$.}
\end{center}
\end{figure}
\begin{figure}[h]
\begin{center}
\fbox{
\includegraphics[scale=0.3]{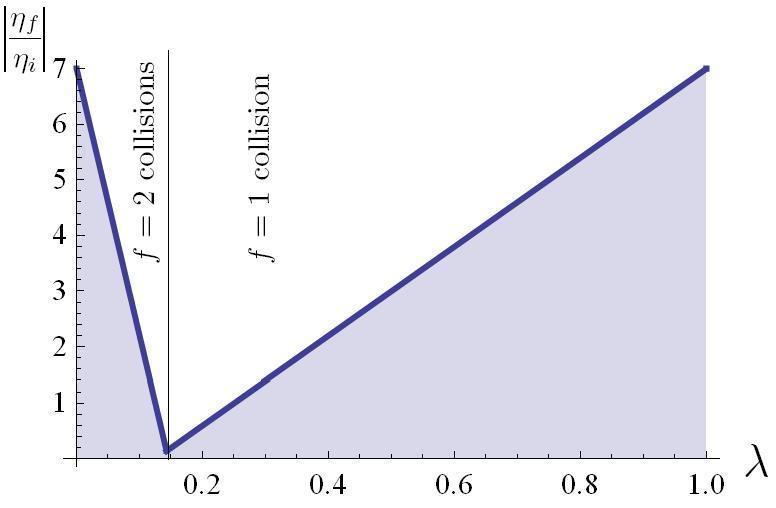}}
\caption{\label{fig:coll}\footnotesize The dependence of $\left|\frac{\eta_{f}}{\eta_{i}}\right|$ on $\lambda$ for the period two diametrical MUPO ($s=2$), where $\eta_{i} \ll 1$ and $\lambda=\frac{d_{1}}{2 \eta_{i} s}$ as described in equation \eqref{eta},}
\end{center}
\end{figure}

We now consider ICs $(\phi_{i},\theta_{i})$ such that the perturbation $\eta_{i}\ll 1$ causes the orbit to initially precess away from the hole. Here, we define $\phi_{i}\in(0,\phi-h^{+})$ to be the angular distance of the IC from the edge of the circular arc at $z=\phi$.
The precessing orbit will first collide with the straight segment after $s n\in \N^{+}$ consecutive $C$-type collisions with the boundary during which time $\theta_{i}$ is constant. $n$ here is the number of MUPO periods the IC will complete just after traversing the available curved distance to $z=\phi$. The first $S$-type collision will occur at $z=\phi+d_{1}$ where $d_{1}= 2 |\eta_{i}| s n -\phi_{i}+ \mathcal{O}(\eta_{i}^{2}) \geq0$. Since the collision occurs on a straight segment of the billiard's boundary, we may `unfold' the billiard using the \textit{image reconstruction trick} \cite{Bun01} (see Figure ~\ref{fig:drivebeltcoll}). The resulting angle of incidence with the curved segment is to leading order $\theta_{f}=\theta_{i}-2d_{1}=\theta_{s,j}+\eta_{f}$. Hence, we conclude that the nonlinear collision process described above causes an orbit initially perturbed by $\eta_{i}$ to have a final incidence angle $\theta_{f}$ which is $\eta_{f}$ away from the original MUPO angle $\theta_{s,j}$ such that $\frac{\eta_{f}}{\eta_{i}}=1- \frac{2 d_{1}}{\eta_{i}}$.
If $2|\eta_{f}|s >d_{1}$ then the orbit will start to precess towards the hole and will eventually escape without jumping over the hole if $2|\eta_{f}|s < \epsilon$. The later is guaranteed for $t$ large enough. The symbolic dynamics characterizing the trajectory of such an IC is therefore given by $C\ldots CSC\ldots C$. If however $2|\eta_{f}|s < d_{1}$, then a second $S$-type collision will occur such that the final angle is now $\theta_{f}=\theta_{s,j}+\eta_{f}= \theta_{s,j}+ 4d_{1}(1-2s) +\eta_{i}(4s-1)$.
The symbolic dynamics sequence of such a collision process is hence given by $C\ldots CSC^{s-1}SC\ldots C$.
The appropriate scaling which linearizes $|\eta_{f}/\eta_{i}|$ is given by $\lambda=\frac{d_{1}}{2 \eta_{i}s}$. Using this we arrive at a small perturbation collision rule given by
\es{\label{eta}\left|\frac{\eta_{f}}{\eta_{i}}\right|=\begin{cases}
-16s^{2}\lambda + s(4+8\lambda)-1, &   0\leq \lambda < \frac{1}{4s -1}, \hspace{0.15in} f=2\\
4\lambda s -1, &  \frac{1}{4s-1}\leq \lambda <1, \hspace{0.15in} f=1. \end{cases}
}
where $f=1,2$ is the corresponding number of curved collisions during the nonlinear collision process.
This is plotted in Figure ~\ref{fig:coll} for the period two diametrical orbit. Notice that when $|\eta_{f}/\eta_{i}|>1$, the return precessing speed of the IC is increased and occurs with higher probability than a decrease. Also, a double ($f=2$) collision scenario is much less probable than a single one ($f=1$).

\begin{figure}[h]
\begin{center}
\fbox{
\includegraphics[scale=0.37]{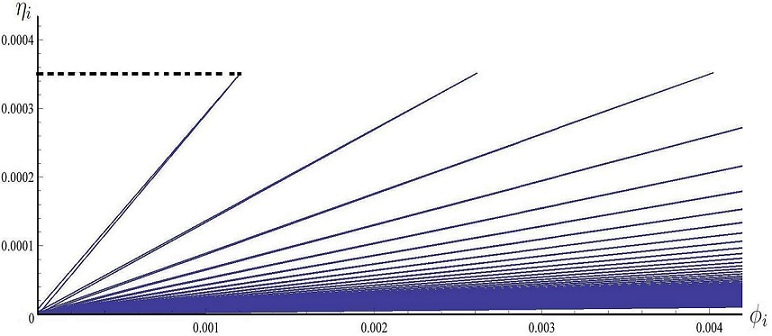}}
\caption{\label{fig:drivspikes}\footnotesize The first $100$ spikes produced by the hyperbolas of equation \eqref{t} for the same parameter values as in Figure ~\ref{fig:drivespikes}.}
\end{center}
\end{figure}

We now formulate the time to escape function:
\es{t(\phi_{i},\theta_{i})&= \left\lceil\frac{\phi_{i}}{2|\eta_{i}|s}\right\rceil s2R\cos\theta_{i}+ \left\lceil\frac{\phi_{1}}{2|\eta_{f}|s}\right\rceil s2R\cos\theta_{f}+ \delta\\
&\approx \frac{R \phi_{i} \cos\theta_{s,j}}{|\eta_{i}|}+ \frac{R \phi_{1} \cos\theta_{s,j}}{|\eta_{f}|},}
to leading order where $\delta= s2R\cos\theta_{s,j} + \mathcal{O}(\eta_{i})$ and we have assumed that both $\eta_{i}$ and $\eta_{f}$ are small.
Notice that again we have obtained the escape time in a continuous time frame from discrete iterations of the billiard map. This is allowed here since the time between collisions is fixed and deviates only slightly during the nonlinear collision process.
We drop the modulus signs from here on, neglect higher order terms in $\eta_{i}$ and $\eta_{f}$, and rearrange to obtain:
\es{\label{t} 0= R \eta_{f} \phi_{i} \cos\theta_{s,j}+ R \eta_{i} \phi_{1} \cos\theta_{s,j} - t \eta_{i}\eta_{f},}
which is a conic section since it is quadratic in both $\phi_{i}$ and $\eta_{i}$ due to the definition of $d_{1}$. Substituting $\eta_{f}$ into \eqref{t} and writing everything in terms of $\phi_{i}$ and $\eta_{i}$, we obtain two hyperbolas for each value of $n$. Each pair of hyperbolas encloses an area which stretches and tilts in a non-overlapping fashion as $n$ is increased. Figure ~\ref{fig:drivspikes} shows how this approximation reproduces the area occupied by the long surviving orbits in the vicinity of the diametrical period two MUPO.

Noticing that the hyperbolas defined by \eqref{t} look like straight lines in Figure ~\ref{fig:drivspikes}, we use the intersection coordinates at the top and bottom of each spike to form lines which maintain the general structure of the enclosed surviving set.
Note that the error in this approximation is only of order $\mathcal{O}(t^{-2})$. This allows for easy integration and summation over the area of each spike in order to obtain $P_{2}(t)$ to leading order. However, care must be taken at large values of $n$ since the spikes will tilt into the hole and hence modifications are necessary. For a step-by-step explanation see \cite{OreDet09}.

Finally, adding all MUPO contributions and expanding for large $t$ we obtain
\es{P_{2}(t)&= \sum_{(s,j)\in \mathcal{S}_{\phi}}  s p \left( \frac{R^{2}(\phi_{1}^{2}+\phi_{2}^{2})\cos^{2}\theta_{s,j}}{4 |\partial Q|t}\right)\\
& \times\Big[1+\frac{(4s-1)}{2s(2s-1)}\ln(4s-1)\Big],
}
where $p$ was defined in the previous subsection and $\phi_{1}$ and $\phi_{2}$ are the allowed ranges on either side of the hole and hence implicitly depend on $s$. Note that this equation is valid for times $t\geq\hat{t}= \frac{8 (s \phi-\pi)}{\epsilon}$ which is larger than \eqref{tt} and can be derived in a similar way.

\subsection{Asymptotic $P(t)$ and numerical simulations}

\begin{figure}[h]
\begin{center}
\fbox{
\includegraphics[scale=0.35]{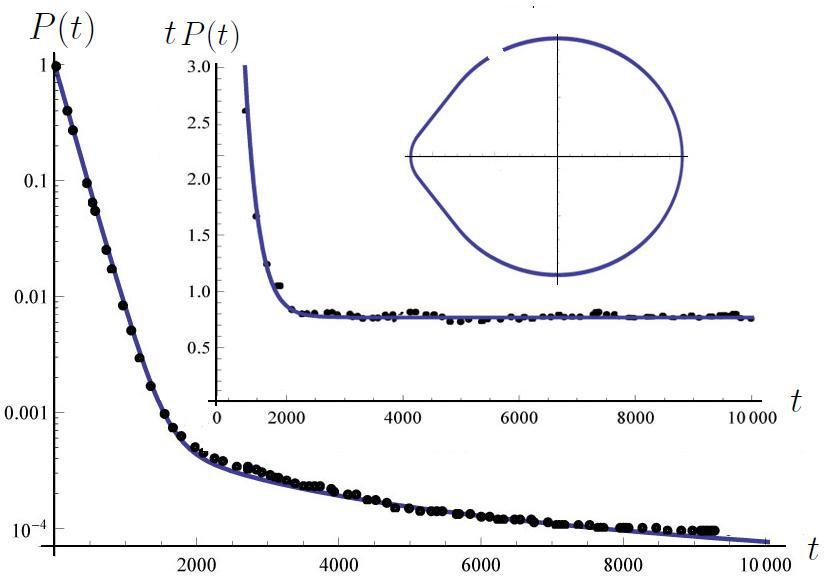}}
\caption{\label{fig:drivept}\footnotesize Log-linear plot of the Survival probability $P(t)$ as a function of time $t$ for the open drivebelt with $\phi=2.5$, $R=1$, $r=0.3$ and a hole of size $\epsilon=0.1$ positioned at $z\in (2.3,2.4)$ as can be seen in the inset. The blue curve is given by the analytic prediction of equation \eqref{Pt} while the empty black circles are from a numerical simulation consisting of $10^{8}$ initial conditions. The inset is a plot of $t P(t)$ showing the agreement with the analytic expressions for the constant $\mathcal{C}$.}
\end{center}
\end{figure}

In the previous subsections we first classified long surviving ICs into two simple families: 1) orbits initially precessing towards the hole, and 2) orbits initially precessing away from the hole. This simplification was achieved by placing the hole as to overlap all MUPO ranges. We then derived conditions defining the sets of the ICs surviving up to time $t$ for each classification and integrated their volume in phase space. Adding the two contributions and taking the limit we obtain for
the survival probability of the open drivebelt with initial conditions chosen according to the
invariant measure on the boundary
\es{\label{pt}
\mathcal{C}=\lim_{t\rightarrow\infty} t P(t) &= \sum_{(s,j)\in \mathcal{S}_{\phi}}  s p \left( \frac{R^{2}(\phi_{1}^{2}+\phi_{2}^{2})\cos^{2}\theta_{s,j}}{4 |\partial Q|}\right)\\
& \times\Big[2+\frac{(4s-1)}{2s(2s-1)}\ln(4s-1)\Big].
}

We now confirm our theoretical predictions through numerical simulations. These are shown in Figure ~\ref{fig:drivept} indicating an excellent agreement with the theory. Other simulations with different $\phi$ and $\epsilon$ have also been performed but are not included here. Moreover, the numerical simulations suggest that the survival probability function exhibits a coexistence of exponential and power law decay as in typical weakly chaotic systems \cite{Vaienti03} given by:
\begin{equation}
\label{Pt}
P(t)=\begin{cases}
\mathrm{irregular}, &   \text{for $t< \hat{t}$}\\
e^{-\gamma t} + \frac{\mathcal{C}}{t}, &  \text{for $t>\hat{t}$,}\end{cases},
\end{equation}
where we have neglected terms of order $\sim t^{-2}$. The `irregular' short-time behavior shows small fluctuations which are a result of geometry dependent short orbits. These become less important if the hole is small. Also, the coefficient of the exponential term is approximately $1$ since for small holes, mixing causes the system to forget its initial state and therefore the probability decays as a Poisson process. 
Finally, equation \eqref{Pt} supports the arguments of \cite{Altman09} in that the chaotic saddle splits into a hyperbolic (exponential) and nonhyperbolic (power-law) part which give separate contributions to $P(t)$.

\subsection{Scaling}

The natural scaling limit for open billiards (or indeed other dynamical systems)
is when $\tau= \epsilon t$ is kept constant while $\epsilon\to 0$ and
$t\to\infty$.  Substituting this into \eqref{e:1} and \eqref{Pt} above we get that
\begin{equation}
 \lim_{\epsilon\rightarrow0} P\left(\tau/\epsilon\right)= e^{-\frac{\tau}{\pi |Q|}},
\end{equation}
since $\mathcal{C}/t=\epsilon\mathcal{C}/\tau$ with $\mathcal {C}$ constant vanishes in this
limit. Hence, the open drivebelt billiard (and also the straight stadium which has a similar
form of the survival probability) satisfies an exponential scaling limit in common with more
chaotic systems of \eqref{e:1}. In other words such scaling approaches ``miss'' the asymptotic
tails and cannot discriminate between strongly chaotic and weakly chaotic systems.
It is worth mentioning that this exponential scaling limit is also intimately connected with
limiting distributions of recurrence times~\cite{HLV, Akaishi09}. An exception to exponential
behaviour is the completely regular circular billiard which shows (numerically) a
nontrivial scaling law \cite{DettBun05}.

\section{Discussion and Conclusions}

We have considered the open drivebelt billiard and generalized the approach of Ref.\cite{OreDet09} in order to obtain exact to leading order expressions for the asymptotic survival probability $P(t)\sim \mathcal{C}/t$.
We have confirmed our predictions through numerical simulations and have obtained a full description of $P(t)$ to leading order.
The generalization of our previous results is based on a small perturbation collision rule derived at the vicinity of smoothly connected circular and straight boundary segments.
Therefore, our approach is now complete and can further be applied to a variety of smooth billiards constructed of conic components and straight lines (for example elliptical stadia).

The quantitative result offered by equation \eqref{pt} and its derivation goes beyond previous investigations of open billiards which were only of qualitative nature in that it provides answers to question like where to place the hole achieve maximum/minimum escape \cite{BunY11}.
This can be done by minimizing or maximizing $\mathcal{C}$ as a function of hole position. For the drivebelt, we find that the maximum $\mathcal{C}$ is attained when the hole is placed right at the edge of the arc i.e. at $z\in(\phi-\epsilon,\phi)$. The minimum is a more complicated function of $\phi$.

Our approach for obtaining the constant $\mathcal{C}$ is however restricted to the case of a hole positioned as to intersect all MUPO ranges and thus splitting the chaotic saddle into non-hyperbolic and hyperbolic parts such that the former occupies a simple region in phase space which can be isolated and integrated over. The case of a hole not intersecting all MUPO ranges remains an open problem although the power-law asymptotic decay $\sim t^{-1}$ is expected to persist. Interestingly, the multiple intermittency of the drivebelt will cause multiple escape paths to exist. For instance, a sticky orbit may exit and then enter into a different sticky mode with some transfer probability before escape occurs. Such scenarios motivate investigations of escape through multiple holes where different mechanisms can be utilized to produce for example \textit{asymmetric transport} as observed in the stadium billiard \cite{OreDett11}. Similarly, one may consider how an increased number of holes intersecting different MUPOs can affect escape through shadowing effects \cite{Paar01}. Such considerations are very relevant in the field of controlling chaos since the escape route through one of the holes may be considerably reduced by other holes.

The results on the asymptotic survival probability in the open drivebelt and the improved understanding of stickiness due to MUPOs presented in this paper are hoped to shed new light on the otherwise not so well understood quantum analogue of stickiness (scarring and anti-scarring).
Similarly, we hope to extend our results in a semiclassical regime
thus offering important ramifications in the context of quantum chaos with experimental applications such as electronic transport through open ballistic nano-structures (quantum dots) \cite{Beenaker}. Finally, by defining conditional recurrence time distributions we expect to be able to manipulate and adapt the multiple intermittency in the drivebelt as to investigate physically interesting scenarios such as achieving high quality factors and directional emission from dielectric cavities \cite{Gmachl98}.

\section*{Acknowledgments}
The authors would like to thank Eduardo Altmann, P\'{e}ter B\'{a}lint and Jens Marklof for illuminating discussions.

\bibliographystyle{unsrt}
\bibliography{drive}

\end{document}